\documentclass{article}
\usepackage{spconf,amsmath,graphicx}
\usepackage[utf8]{inputenc}
\usepackage{hyperref}
\usepackage{subcaption}
\usepackage{multirow}
\usepackage{comment}

\usepackage{glossaries}
\newacronym{asic}{ASIC}{Application-Specific Integrated Circuit}
\newacronym{awgn}{AWGN}{Additive White Gaussian Noise}

\newacronym{bm3d}{BM3D}{Block-Matching 3D}

\newacronym{cnn}{CNN}{Convolutional Neural Network}
\newacronym{crt}{CRT}{Cathode Ray Tube}
\newacronym{cpu}{CPU}{Central Processing Unit}
\newacronym{cr}{CR}{Character Recognition}

\newacronym{dvi}{DVI}{Digital Visual Interface}
\newacronym{dncnn}{DnCNN}{Denoising Convolutional Neural Network}
\newacronym{dnn}{DNN}{Deep Neural Network}
\newacronym{dp}{DP}{Display Port}
\newacronym{dl}{DL}{Deep Learning}

\newacronym{em}{EM}{Electro Magnetic}
\newacronym{emsec}{EMSEC}{Emission Security}
\newacronym{xai}{XAI}{eXplainable Articial Intelligence}

\newacronym{fcn}{FCN}{Fully Convolutional Network}
\newacronym{fc}{FC}{Fully Connected}
\newacronym{fpga}{FPGA}{Field Programmable Gate Array}
\newacronym{fpn}{FPN}{Fixed Pattern Noise}

\newacronym{gpu}{GPU}{Graphics Processing Unit}
\newacronym{gan}{GAN}{Generative Adversarial Networks}

\newacronym{hdmi}{HDMI}{High-Definition Multimedia Interface}

\newacronym{ipe}{IPE}{Information processing equipment}
\newacronym{ilsvrc}{ILSVRC}{ImageNet Large Scale Visual Recognition Competition}

\newacronym{lcd}{LCD}{Liquid Crystal Display}
\newacronym{lvds}{LVDS}{Low-Voltage Differential Signaling}

\newacronym{mse}{MSE}{Mean Square Error}
\newacronym{mae}{MAE}{Mean Absolute Error}

\newacronym{ocr}{OCR}{Optical Character Recognition}

\newacronym{psnr}{PSNR}{Peak Signal to Noise Ratio}
\newacronym{roi}{RoI}{Region of Interest}
\newacronym{rois}{RoIs}{Regions of Interest}
\newacronym{rf}{RF}{Radio Frequency}
\newacronym{rmse}{RMSE}{Root Mean Square Error}
\newacronym{rpn}{RPN}{Region Proposal Network}

\newacronym{sdr}{SDR}{Software-Defined Radio}
\newacronym{snr}{SNR}{Signal to Noise Ratio}
\newacronym{ssim}{SSIM}{Structure Similarity}

\newacronym{vga}{VGA}{Video Graphics Array}
\title{Electro-Magnetic Side-Channel Attack\\Through Learned Denoising and Classification}

\name{\begin{tabular}{c}Florian Lemarchand $^{\star}$, Cyril Marlin $^{\mathsection}$, Florent Montreuil $^{\mathsection}$\\ Erwan Nogues $^{\star \mathsection}$, Maxime Pelcat $^{\star}$\end{tabular}}

\address{$^{\star}$ Univ. Rennes, INSA Rennes, IETR - UMR CNRS 6164 \\
	    $^{\mathsection}$ DGA-MI, Bruz}
\begin{document}
\ninept
\maketitle
\begin{abstract}
This paper proposes an upgraded \gls{em} side-channel attack that automatically reconstructs the intercepted data.
A novel system is introduced, running in parallel with leakage signal interception and catching compromising data in real-time.
Based on deep learning and \gls{cr} the proposed system retrieves more than 57\% of characters present in intercepted signals regardless of signal type: analog or digital.
The approach is also extended to a protection system that triggers an alarm if the system is compromised, demonstrating a success rate over 95\%.
Based on \gls{sdr} and \gls{gpu} architectures, this solution can be easily deployed onto existing information systems where information shall be kept secret.
\end{abstract}

% -------------------------------------------------------------------------

% -------------------------------------------------------------------------
\glsresetall
\section{Introduction}
\label{sec:intro}
All electronic devices produce \gls{em} emanations that not only interfere with radio devices but also compromise the data handled by the information system. A third party may perform a side-channel analysis and recover the original information, hence compromising the system privacy. While pioneering work of the domain focused on analog signals~\cite{van_eck_electromagnetic_1985}, recent studies extend the eavesdropping exploit using an \gls{em} side-channel attack to digital signals and embedded circuits~\cite{kuhn_compromising_2013}. 
The attacker's profile is also taking on a new dimension with the increased performance of \gls{sdr}. With recent advances in radio equipment, an attacker can leverage on advanced signal processing to further stretch the limits of the side-channel attack using \gls{em} emanations~\cite{genkin_synesthesia:_2018}. The fast evolution of deep neural networks, an attacker can extract patterns or even the full structured content of the intercepted data with a high degree of confidence and a limited execution time.

% Claims
%In this paper, we propose a deep learning-based method to denoise and interpret an intercepted signal. 
In this paper, a learning-based method is proposed with the specialization of Mask R-CNN~\cite{he_mask_2017} as a denoiser and classifier. 
%and dramatically increase the performance of conventional image processing techniques. 
A complete system is demonstrated, embedding \gls{sdr} and deep-learning, that detects and recovers leaked information at a distance of several tens of meters. It provides an automated solution where the data is interpreted directly. The solution is compared to other system setups.

The paper is organized as follows. Section~\ref{sec:related_work} presents existing methods to recover information from \gls{em} emanations.
Section~\ref{sec:prop_meth} describes the proposed method for automatic character retrieval.
Experimental results and detailed performances are exposed in Section~\ref{sec:test_results}. Section~\ref{sec:conclu} concludes the paper. 
% -------------------------------------------------------------------------
\section{Related Work}\label{sec:related_work}
% Paragraph Tempest (Signaux Parasites)
This research focuses on two areas: \gls{em} side channel attacks on information systems and learning-based techniques that can recover information from noisy environments. 

Van Eck \textit{et al.}~\cite{van_eck_electromagnetic_1985} published the first technical reports revealing how involuntary emissions originating from electronics devices can be exploited to compromise data. While the original work of the domain targeted \gls{crt} screens and analog signals, Kuhn \textit{et al.}~\cite{kuhn_compromising_2013} propose to use side-channel attacks to extract confidential data from \glspl{lcd}, targeting digital data. Subsequently, other types of systems have been attacked. Vuagnoux \textit{et al.}~\cite{vuagnoux_compromising_2009} extend the principle of \gls{em} side-channel attack to capture data from keyboards and, in their recent work, Hayashi \textit{et al.} present interception methods based on \gls{sdr} targeting laptops, tablets\cite{hayashi_threat_2014} and smartphones~\cite{hayashi_remote_2017}. The use of \gls{sdr} increases the surface of attack from military organizations to hackers. It also opens up new post-processing opportunities that improve attack characteristics. %Using the \gls{sdr} equipments raises up the risk associated to \gls{em} side-channel attack.
De Meulemeester \textit{et al.}~\cite{de_meulemeester_synchronization_2018} leverage on \gls{sdr} to enhance the performance of the attack and automatically find the structure of the captured data. When the intercepted emanation is originally 2D, retrieving the synchronization parameters of the targeted information system, the captured \gls{em} signal can be transformed from a vector to an image, reconstructing the 2-dimensional sensitive visual information. That step is called the \textit{rastering}.

\begin{figure*}
    \centering
    \includegraphics[width=0.9\linewidth]{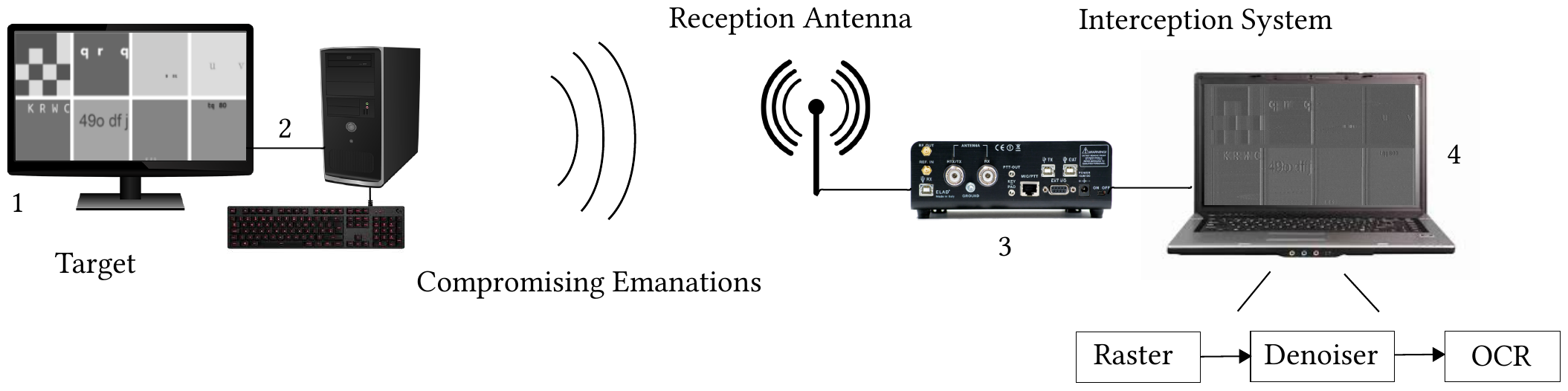}
\caption{Experimental setup: the attacked system includes an eavesdropped screen (1) displaying sensitive information. It is connected to an information system (2). An interception chain including an \gls{sdr} receiver (3) sends samples to a host computer (4) that implements signal processing including a deep learning denoiser and \gls{cr}.}
  \label{fig:complete_system} 
\end{figure*}

When retrieving visual information from an \gls{em} signal, an important part of the original information is lost through the leakage and interception process. This loss leads to a drop of the \gls{snr} and a deterioration of spatial coherence into the reconstructed samples in the case of image data. Hence, denoising methods are needed.
Image denoising by signal processing techniques has been extensively studied since it is an important step in many computer vision applications. 
BM3D~\cite{dabov_image_2007}, proposed by Dabov \textit{et al.}, is a the state-of-the-art methods for \gls{awgn} removal using non-learned processing. BM3D uses thresholding and Wiener filtering into the transform domain. It is used in the experiments of Section~\ref{sec:test_results}.

Deep learning algorithms have recently stood out from the crowd for solving many signal processing problems. These trained models have an extreme ability to fit complex problems. Recent \gls{gpu} architectures have been optimized to support deep learning workloads and have fostered ever deeper networks, mining structured information from data and providing results where classical algorithms fail. The spread of deep learning has occurred in the domain of image denoising and several models initially developed for other applications have been turned into denoisers. \gls{dncnn}~\cite{zhang_beyond_2017} is a \gls{cnn} designed to blindly remove \gls{awgn}, without prior knowledge on noise level. Others techniques such as denoising autoencoders~\cite{vincent_stacked_2010, ronneberger_u-net:_2015} are able to denoise images without restriction on the type of noise. Autoencoders algorithms learn to map their input to a latent space (encoding) and project back the latent representation to the input space (decoding). Autoencoders learn a denoising model by minimizing a loss function which evaluates the difference between the autoencoder output and the reference. Advanced methods, such as Noise2Noise~\cite{lehtinen_noise2noise:_2018}, infer denoising strategies without any clean input reference data. Noise2Noise algorithm learns a representation of the noise by looking only at noisy samples. 

Learning-based models perform well in various denoising but with strong hypothesis regarding the distribution of the noise to be withdrawn. \gls{awgn} assumption is often used.
In the considered problem, certain components of the noise are non-randomly distributed and have a spatial coherence (between pixels). Additionally, information is damaged (partially lost and spread over several pixels) by the interception/rastering process. None of the previously exposed methods is tailored for such noise and distortion natures, calling for a novel experimental setup.

Conventional approaches exist to protect devices from eavesdropping. Such approaches appear under different code names such as TEMPEST~\cite{national_security_agency_nacsim_1982} or \gls{emsec} and consist of shielding devices \cite{kuhn_compromising_2013} to nullify the emanations, or using fonts that minimize the \gls{em} emanations~\cite{aucsmith_soft_1998}. However, these approaches are either costly solutions or technically hard to use in practice especially when it comes to ensure the data privacy throughout the life-cycle of a complex information system. The next section details the proposed method to enhance the \gls{em} side-channel attack.
%++++++++++++++++++++++++++++++++++++++++++++++++++++++++++++++++++++++++++++
\section{Proposed Side-Channel Attack}\label{sec:prop_meth}

\subsection{System Description}\label{subsec:sys_descript}

Figure~\ref{fig:complete_system} shows the proposed end-to-end solution. The method automatically reconstructs leaked visual information from compromising emanations. The setup is composed of two main elements. At first the antenna and \gls{sdr} processing capture in the \gls{rf} domain the leaked information originating from the displayed video. Then, the demodulated signal is processed by the host computer, recovering a noisy version of the original image \cite{kuhn_compromising_2013} leaving room for advanced image processing techniques.
On top of proposing an end-to-end solution from the capture to the data itself, the method uses a learning-based approach. It exploits the capturing compromising signals and recognized automatically the leaked data. A first step based on a Mask R-CNN (Mask R-CNN) architecture embeds the following: denoising, segmentation, character detection/localization, and character recognition. A second step post-processes the Mask R-CNN output. A Hough transform is done for text line detection and a Bitap algorithm~\cite{myers_fast_1999} is applied to approximate match information. Thi setup detects several forms of compromising emanations (analog or digital) and automatically triggers an alarm if critical information is leaking. Next sections detail how the method is trained and integrated.

% On parle de l'interception, du débruitage puis notre contribution : denoiser + ocr
\subsection{Training Dataset Construction}\label{subsec:dataset_const}

A substantial effort has been made on building a process that semi-automatically generates and labels datasets for supervised training. Each sample image is made up of a uniform background on which varied characters are printed. Using that process, an open data corpus of 123.610 labeled samples, specific to the problem at hand, has been created to further be used as training, validation and test datasets. This dataset is available online~\footnote{\url{https://github.com/opendenoising/interception_dataset}} to train denoiser architectures in difficult conditions.

The proposed setup, to be trained, denoises the intercepted sample images and extracts their content, i.e. the detected characters and their positions. The input space that should be covered by the training dataset is large and three main types of interception variability can be observed. Firstly, interception induces an important loss of the information originally existing in the intercepted data. The noise level is directly linked to the distance between the antenna and the target. Several noise levels are generated by adding \gls{rf} attenuation after the antenna.
That loss itself causes inconsistencies in the rasterizing stage. Secondly, \gls{em} emanations can come from different sources, using different technologies, implying in turn different intercepted samples for the same reference image. The dataset covers \gls{vga}, \gls{dp}-to-\gls{dvi} and \gls{hdmi} cables and connectors. Besides this unwanted variability, a synthetic third type of variability is introduced to solve the character retrieval. Many different characters are introduced in the corpus to be displayed on the attacked screen. They range from 11 to 70 points in size and they are both digits and letters, and letters are both upper and lower cases. Varied fonts, character colors and background colors, as well as varied character positions in the sample are used. Considering these different sources of variability, the dataset is built trying to get an equi-representation of the different interception conditions. 

\begin{figure}
    \centering
    \includegraphics[width=\linewidth]{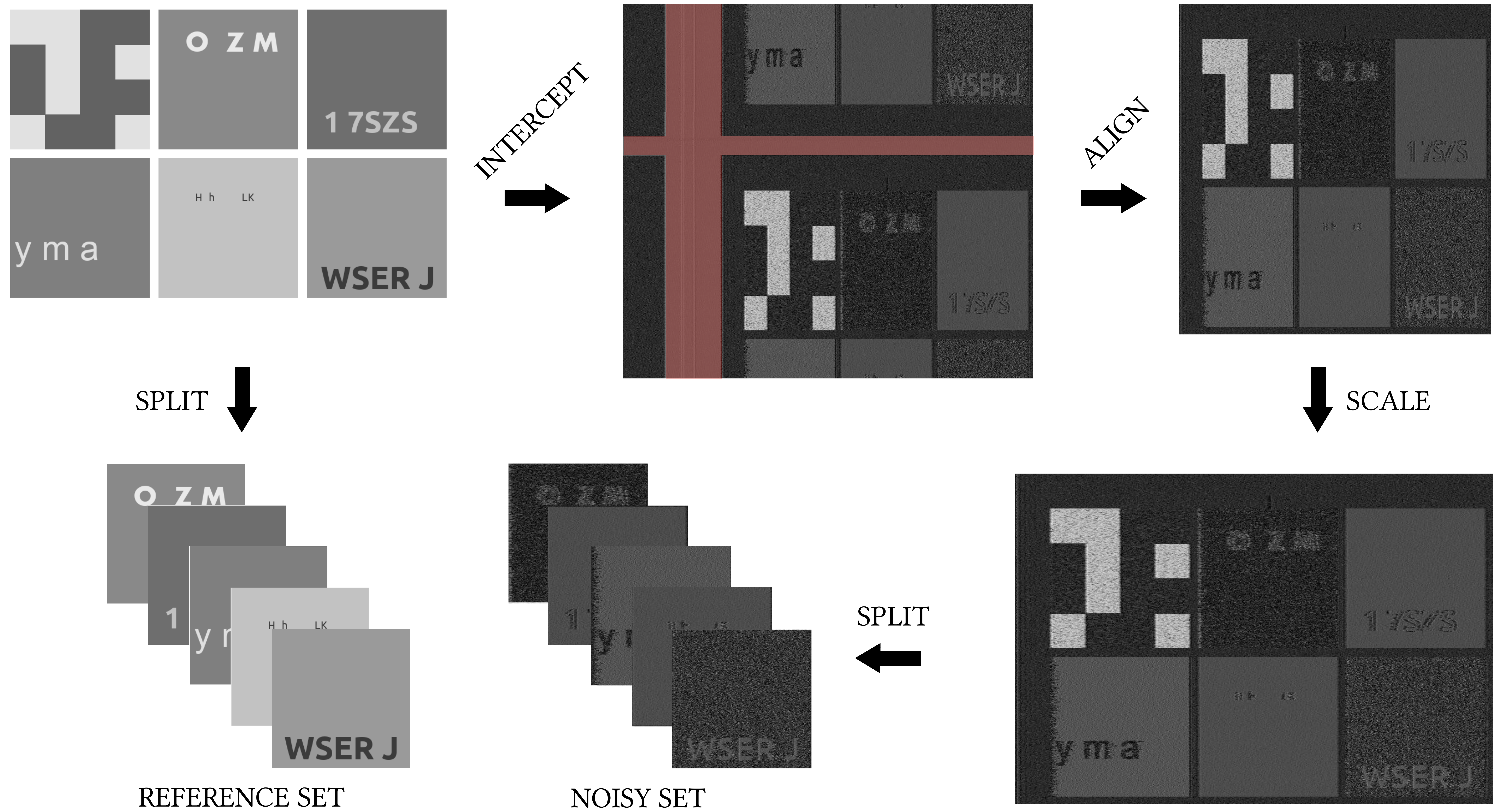}
    \caption{A reference sample is displayed on the target screen (top-left). The interception module outputs uncalibrated samples. Vertical and horizontal porchs (red) helps alignment and porch withdrawal (top-right). Samples are rescaled and split into patches to obtain the same layout than the reference set.}
  \label{fig:make_dataset}
\end{figure}

The choice has been made to display on the target screen a sample containing patches of size $256\times256$ pixels (top-left image of Figure~\ref{fig:make_dataset}). For building the dataset, having multiple patches speeds the process up because smaller samples can be derived from a single screen interception and more variability can be introduced in the dataset. The main challenge when creating the dataset lies in the samples acquisition itself. Indeed, once intercepted, the samples are not directly usable. The interception process outputs samples such as the one of Figure~\ref{fig:make_dataset} (middle-top) where intercepted characters are not aligned (temporally and spatially) with respective reference samples. An automated method is introduced that uses the porches, artificially colored in red in Figure~\ref{fig:make_dataset} (middle-top), to align spatially samples. Porches are detected using brute-force search of large horizontal and vertical gradients (to find vertical and horizontal porches, respectively). A validation step ensures the temporal alignment, based on the insertion of a QRCode in the upper-left patch. If the QRCode is similar between the reference and the intercepted image, the image patches are introduced in the dataset.

Data augmentation~\cite{mikolajczyk_data_2018} is used to enhance the dataset coverage area. It is done onto patches to add variability into the dataset and reinforce its learning capacity. Conventional methods are applied to raw samples to linearly transform them (Gaussian and median blur, salt and pepper noise, color inversion and contrast normalization). 
%Examples of applied transforms include Gaussian and median blurs, salt and pepper noise, color inversion, or contrast normalization. 
 
\subsection{Implemented Solution to Catch Compromising Data}\label{subsec:implem_sol}

In order to automate the interception of compromising data, the Mask R-CNN has been turned into a denoiser and classifier. The implementation is based on the one proposed by W. Abdulla~\footnote{~~\url{https://github.com/matterport/Mask\_RCNN}}. Other learning-based and classical signal processing methods, discussed in Section~\ref{subsec:perf_comp}, are also implemented to assess the quality of the proposed framework. Mask R-CNN is a framework adapted from the previous Faster R-CNN~\cite{ren_faster_2017}. The network consists of two stages. The first stage, also known as \textit{backbone} network, is a \emph{ResNet101} convolutional network~\cite{he_deep_2016} extracting features out of the input samples. Based on the extracted features, a \gls{rpn} proposes \glspl{roi}. \glspl{roi} are regions in the sample where information deserves greater attention. The second stage, called \textit{head} network, classifies the content and returns bounding box coordinates for each of the \glspl{roi}. The main difference between Faster R-CNN and Mask R-CNN lies in an additional \gls{fcn} branch~\cite{shelhamer_fully_2017} running in parallel with the classification and extracting a binary mask for each \gls{roi} to provide a more accurate localization of the object of interest. 

 Mask R-CNN is not originally designed to be used for denoising but rather for instance segmentation. However, it fits well the targeted problem. Indeed, the problem is similar to a segmentation where signal has to be separated from noise. As a consequence, when properly feeding a trained Mask R-CNN network with noisy samples containing characters, one obtains lists of labels (i.e. characters recognition), as well as their bounding boxes (characters localization) and binary masks representing the content of the original \emph{clean} sample. The 
setup of the classification branch allows to be language-independent and to add classes other than characters.
 
\begin{figure}
  \centering
    \includegraphics[width=\linewidth]{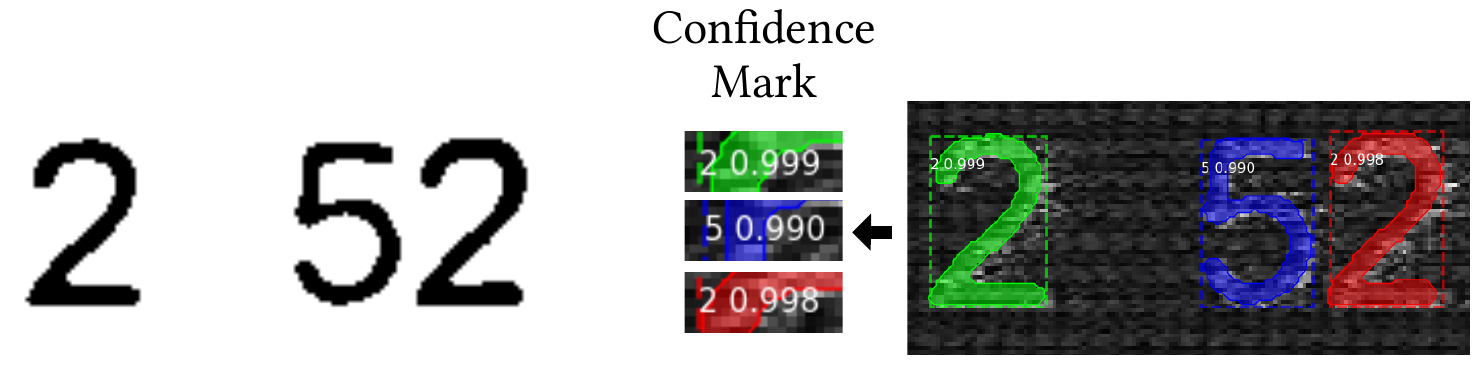}
  \caption{The output of Mask R-CNN may be used in two ways. The segmentation can be drawn (left) and further processed by an \gls{ocr}, or the Mask R-CNN classifier can directly infer the sample content (right) and propose some display and confidence information.}
  \label{fig:output_mrcnn}
\end{figure}

Two strategies can be employed to exploit Mask R-CNN components for the problem. The first idea is to draw the output masks of Mask R-CNN segmentation (Figure~\ref{fig:output_mrcnn} left-hand side) and request an \gls{ocr} to retrieve characters from the masks. A second possibility is to make use of the classification faculty of Mask R-CNN (Figure~\ref{fig:output_mrcnn} right-hand side) and obtain a list of labels without using an \gls{ocr} engine. The second method using the classifier of Mask R-CNN proves to be better in practice, as shown in Section~\ref{subsec:perf_comp}.

The training strategy is to initialize the training process using pre-trained weights~\cite{hebert_exploring_2018} for the MS COCO~\cite{lin_microsoft_2014} dataset. First, the weights of the \textit{backbone} are frozen and the \textit{head} is trained to adapt to the application. Then, the weights of the \textit{backbone} are relaxed and both \textit{backbone} and \textit{head} are trained together until convergence. This process is done to ensure the convergence and speed up training.

\begin{figure}
  \centering
    \begin{subfigure}[b]{0.3\linewidth}
    \includegraphics[width=\linewidth]{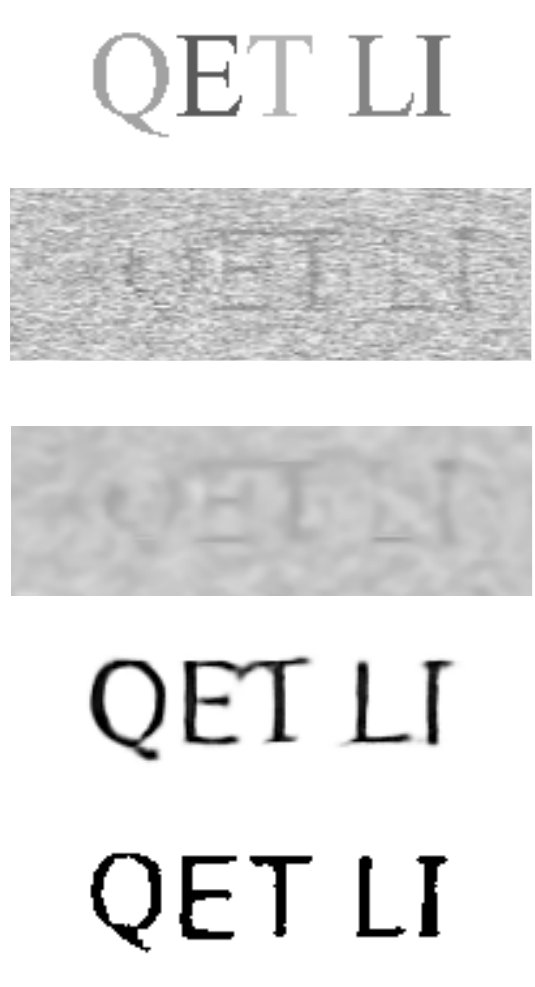}
  \end{subfigure}
    \begin{subfigure}[b]{0.3\linewidth}
    \includegraphics[width=\linewidth]{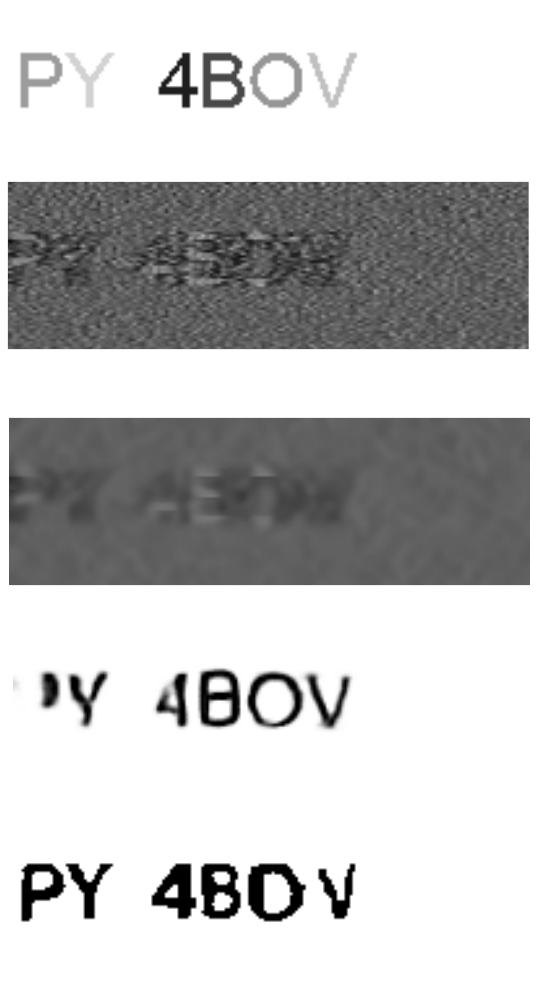}
  \end{subfigure}
    \begin{subfigure}[b]{0.3\linewidth}
    \includegraphics[width=\linewidth]{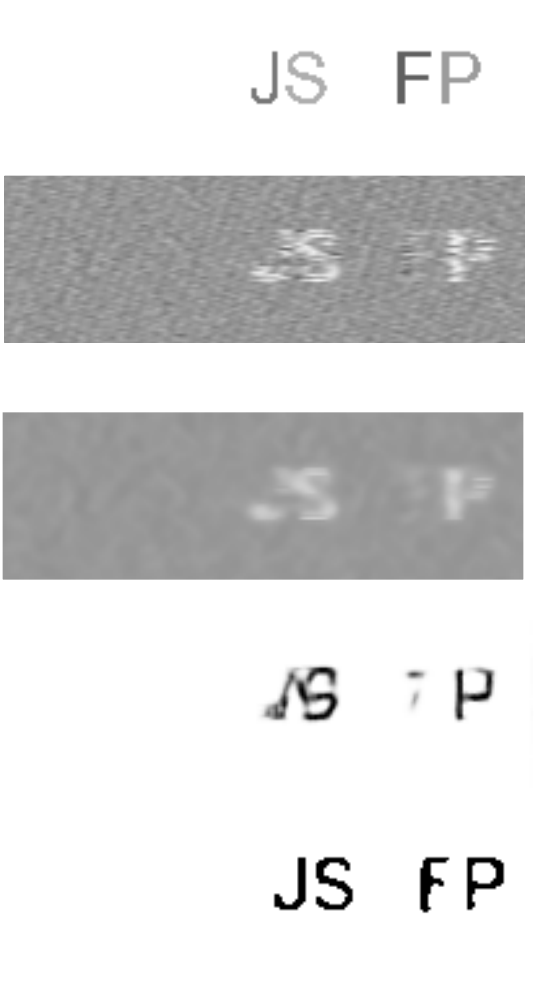}
  \end{subfigure}
  \caption{Three samples (left, middle, right) displayed at different stages of the interception/denoising pipeline. From top to bottom: the reference patch displayed on the screen; the patch after rasterization (raw patch); the patches denoised with \gls{bm3d}, autoencoder and Mask R-CNN.}
  \label{fig:pipeline}
\end{figure}
%++++++++++++++++++++++++++++++++++++++++++++++++++++++++++++++++++++++++++++
\section{Experimental Results}\label{sec:test_results}
\subsection{Experimental Setup}\label{subsec:test_setup}

The experimental setup is defined as follows: the eavesdropped display is 10 meters away from the interception antenna. A \gls{rf} attenuator is inserted after the antenna. It ranges from 0 dB to 24 dB to generate a wide range noise values and simulated higher interception radius as shown in the examples~\ref{fig:output_mrcnn} . Compromising emanations are issued either by a \gls{vga} display, a \gls{dp}-to-\gls{dvi} cable or an \gls{hdmi} connector. The interception system is depicted in Figure~\ref{fig:complete_system}: the antenna is bilog, the \gls{sdr} device automatically recovering parameters~\cite{de_meulemeester_synchronization_2018} is an Ettus X310 receiving with a 100 MHz bandwidth to recover the compromised information with a fine granularity \cite{kuhn_compromising_2013}. The host computer running post-processing has a linux operating system, an Intel\textregistered Xeon\textregistered W-2125 \gls{cpu} and an Nvidia GTX 1080 Ti \gls{gpu}. The host computer rasters the compromising data using the \gls{cpu} while the proposed learning-based denoiser/classifier runs also on the \gls{gpu}. 

%% a compleéter et améliorer

\subsection{Performance Comparison Between Data Catchers}\label{subsec:perf_comp}

The purpose of the exposed method is to analyze compromising emanations. Once a signal is detected and rasterized, intercepted emanations should be classified into compromising or not. Figure~\ref{fig:pipeline} illustrates the outputs of different implemented denoisers. More examples are available at~\footnote{\url{https://github.com/opendenoising/extension}}. It is proposed to assess the data leak according to the ability of a model to retrieve original information. A ratio between the number of characters that a method correctly classifies from an intercepted sample, and the true number of characters in the corresponding \emph{clean} reference is used as a metric. 

The quality assessment method is the following.
First, a sample containing a large number of characters is pseudo-randomly generated (similar to dataset construction).
The sample is displayed on the eavesdropped screen and \gls{em} emanations are intercepted. The proposed denoising/retrieval is applied and the obtained results are compared to the reference sample.
The method using Mask R-CNN produces directly a list of retrieved characters.
Other methods, implemented to compare the efficiency of the proposal, use denoising in combination with the Tesseract~\cite{smith_overview_2007} \gls{ocr}.
Tesseract is a well performing \gls{ocr} engine, retrieving characters from images.
It produces a list of characters retrieved from a denoised sample.
As the output of Tesseract is of the same type as the output of Mask R-CNN classification, metrics can be extracted to fairly compare methods. 

An end-to-end evaluation is used measuring the quality of characters classification.  A \textit{F-score} classically used to evaluate classification model is computed using $precision$ and $recall$.
$precision$ is the number of true positives divided by the number of all positives.
$recall$ is the number of true positives divided by the number of relevant samples, the set of relevant samples being the union of true positives and false negatives.
For simplification and not use an alignment process, a true positive is chosen here to be the recognition of a character truly existing in the reference sample.

\begin{table}
    \centering
        \begin{tabular}{ |c|c|c|c|c| }
            \hline
            Denoiser & \gls{ocr} & F-Score & precision & recall \\
            %\hhline{|=|=|=|}
            \hline\hline
            Raw   & \multirow{6}{*}{Tesseract}    & 0.04 &0.20 &  0.02\\
            BM3D &   & 0.13  &0.22 & 0.09 \\
            Noise2Noise &   & 0.17 &0.25 & 0.12  \\
            AutoEncoder &   &0.24 & 0.55& 0.15 \\
            RaGAN  &   & 0.24&0.42 & 0.18 \\
            UNet  &    & 0.35& 0.62&  0.25\\
            Mask R-CNN  &    & 0.55& \textbf{0.82} & 0.42 \\
            \cline{2-2}
            Mask R-CNN &  Mask R-CNN  & \textbf{0.68} & \textbf{0.81} & \textbf{0.57} \\
            \hline
        \end{tabular}
    \caption{Character recognition performance for several data catchers using either denoising and Tesseract, or Mask R-CNN (Mask R-CNN) classification. Mask R-CNN classifier outperforms others methods with a $0.68$ \textit{F-score} on the test set.} 
    \label{tab:res}
\end{table}

Table~\ref{tab:res} presents the results of different data catchers on a test set of 12563 patches.
All denoising methods are tested using Tesseract, and compared to Mask R-CNN classification used as \gls{ocr}.
Tesseract is first applied to raw (non-denoised) samples as a point of reference. \gls{bm3d} is the only classical denoising solution tested. \emph{Noise2Noise}, \emph{AutoEncoder}, \emph{RaGAN} and \emph{UNet} are different deep learning networks configured as denoisers. As shown in Table~\ref{tab:res}, Mask R-CNN classification outperforms all other methods. The version of Mask R-CNN using its own classifier is better than the Tesseract \gls{ocr} engine applied on Mask R-CNN segmentation mask output. It is also interesting to look at precision and recall scores that compose the \textit{F-score}. Both Mask R-CNN methods perform better than other methods for the two indices. Precision is almost the same for both methods, meaning that they both present the same ratio of good decision. The difference lies in the recall score. The $0.42$ recall score of the version using Tesseract is lower than the $0.57$ score of the method using its own classifier, indicating that the latter version miss less characters. The main advantage of the Mask R-CNN is that the processing tasks to solve the final aim of textual information recovery are jointly optimized. 

Another key performance indicator of learning-based algorithms is inference time (Table~\ref{tab:res_time}). The proposed implementation using Mask R-CNN infers results from an input sample of resolution $1200\times1900$ in $4.04$s in average. This inference time, although lower than \gls{bm3d} latency, is admittedly higher than other neural networks and hardly real-time. Nevertheless, the inference time of Mask R-CNN includes all the denoising/\gls{ocr} process and provides a largely better retrieval score. In the context of a continuous listening of \gls{em} emanations, it provides an acceptable trade-off between processing time and interception performance. The optimization of the inference time could be considered as a future work with the recent advances in accelerating neural network inference~\cite{zhang_caffeine:_2016,ferrari_amc:_2018}. 

\begin{table}
    \centering
        \begin{tabular}{ |c|c|c| }
            \hline
            Denoiser & \gls{ocr} & Inference Timing (s) \\
            \hline\hline
            Raw   & \multirow{4}{*}{Tesseract}    & 0.19   \\
            BM3D &     & 21.8\\
            Autoencoder &     & 1.15\\
            Mask R-CNN  &     &4.22\\
            \cline{2-2}
            Mask R-CNN &  Mask R-CNN  & 4.04\\
            \hline
        \end{tabular}
    \caption{Inference time for several data catchers using Tesseract or Mask R-CNN classification as \gls{ocr}. Input resolution is $1200\times1900$ and it is processed using a split in $28$ patches. Mask R-CNN classifier is slower than the autoencoder but still faster than BM3D.}
    \label{tab:res_time}
\end{table}

%\gls{bm3d} has been used here as a reference mark to deep learning based solutions since it uses classical image processing techniques. The scores of \gls{bm3d} show that it is not efficient for the considered application. The poor performances of \gls{bm3d} are due to the complex nature of the addressed noise, not only made of a Gaussian white noise. 

% -------------------------------------------------------------------------
\section{Conclusions}\label{sec:conclu}
Handling data while ensuring trust and privacy is challenging for information system designers. This paper presents how the attack surface can be enlarged with the introduction of deep learning in an \gls{em} side-channel attack. The proposed method uses Mask R-CNN as denoiser and it automatically recovers more than $57\%$ of a leaked information for a wide range of interception distances. The proposal is software-based, and runs on the host computer of an off-the-shelf \gls{sdr} platform. % Hence, we propose to use the method to monitor a deployed information system in real-time and detect any defect appearing on the system.
% -------------------------------------------------------------------------

\bibliographystyle{IEEEbib}
\bibliography{references}
\end{document}